\documentclass[10pt,conference]{IEEEtran}

\usepackage[dvips]{graphicx}  
\usepackage{subfigure}

\begin{document}
\def\picture #1 by #2 (#3 scaled #4 #5){\leavevmode
 \vbox to #2{
   \hrule width #1 height 0pt depth 0pt
   \vfill
   \special{picture #3 scaled #4 #5}}}


\newcommand{\sinc}[1]{\mbox{sinc}\left({#1}\right)}
\newcommand{\erf}[1]{\mbox{erf}\left({#1}\right)}
\newcommand{\erfc}[1]{\mbox{erfc}\left({#1}\right)}
\newcommand{\corr}[1]{\mbox{corr}\left({#1}\right)}
\newcommand{\cov}[1]{\mbox{cov}\left({#1}\right)}
\newcommand{\var}[1]{\mbox{var}\left({#1}\right)}
\newcommand{\sgn}[1]{\mbox{sgn}\left({#1}\right)}
\newcommand{\trace}[1]{\mbox{Tr}\left({#1}\right)}

\newcommand{\randq}{Q}
\newcommand{\randz}{Z}
\newcommand{\randv}{V}
\newcommand{\randh}{H}
\newcommand{\randy}{Y}
\newcommand{\randn}{N}
\newcommand{\randx}{X}
\newcommand{\rands}{S}
\newcommand{\randw}{W}
\newcommand{\vh}{\vec{h}}
\newcommand{\vy}{\vec{y}}
\newcommand{\vx}{\vec{x}}
\newcommand{\rvz}{\vec{Z}}
\newcommand{\rvq}{\vec{Q}}
\newcommand{\rvy}{\vec{Y}}
\newcommand{\rvx}{\vec{X}}
\newcommand{\rvn}{\vec{N}}
\newcommand{\rvh}{\vec{H}}
\newcommand{\rvs}{\vec{S}}
\newcommand{\rvb}{\vec{B}}
\newcommand{\rvv}{\vec{V}}
\newcommand{\rvtz}{{\bar{Z}}}
\newcommand{\rvtq}{{\bar{Q}}}
\newcommand{\vtm}{\bar{m}}
\newcommand{\mtrxh}{{\bf h}}
\newcommand{\mtrxx}{{\bf x}}
\newcommand{\mtrxg}{{\bf g}}
\newcommand{\rmh}{{\bf H}}
\newcommand{\rmx}{{\bf X}}
\newcommand{\rmg}{{\bf G}}
\newcommand{\rmy}{{\bf Y}}
\newcommand{\rmn}{{\bf N}}
\newcommand{\rmw}{{\bf W}}

\title{Antenna Array Geometry and Coding Performance}

\author{\authorblockN{Weijun Zhu, Heechoon Lee, Daniel Liu, and Michael P. Fitz}
\authorblockA{UnWiReD Laboratory\\
University of California Los Angeles \\
Email: unwired@ee.ucla.edu}
\vspace{-0.4in}
}
%

\maketitle

\begin{abstract}
This paper provides details about experiments in realistic, urban, and frequency flat channels with
space-time coding that specifically examines the impact of the number of receive antennas
and the design criteria for code selection on the performance.  Also the performance characteristics are
examined of the coded modulations in the presence of finite size array geometries.  This paper gives some
insight into which of the theories are most useful in realistic deployments.
\end{abstract}

\section{Introduction}
Over the past several years, there has been a great deal of research to
improve performance of wireless communications in fading environments by
exploiting transmitter and/or receiver diversity. The pioneering work by
Telatar \cite{Tel:95}, Foschini and Gans \cite{FG:98} showed that
multiple antennas in a wireless communication system can greatly improve
performance.  For
$L_t$ transmit antennas and $L_r$ receive antennas in Rayleigh fading, it was shown that with
spatial independence there are essentially $L_tL_r$ levels of diversity available and there are
$\min\left(L_t, L_r\right)$ independent parallel channels that could be established. These
information theoretic studies spawned two lines of work; one where the
number of independent channels is large \cite{Fos:96} and one where the
number of independent channels is small \cite{TSC:98,Gea:96}. With eight
years of intensive engineering research and development effort after these
insights, MAR techniques are making a significant impact on how
wireless services are provided.  Examples include the nascent 802.11n
standard and 3G and 4G mobile telecommunications systems.  The efforts in
this area have reached the point where researchers are calling the area
mature.
\subsection{Open Problems in Space-Time Signaling}
The open problems in
MAR communications relate to situations where more sophisticated and
detailed aspects of communication systems need to be modeled and
understood.  For example performance is not easily understood in channel
models that are not well modeled as Gaussian/Rayleigh, or where the
scattering is not rich or isotropic, or where time-varying parameters, or
system non-idealities impact system performance.  These problems are not
well addressed by simulation or analysis as the sophistication of the
problem prohibits analysis in most cases and the utility of simulation is
limited to the accuracy of the models used for simulation.

This paper examines a small subset of the open issues in the
literature and reports on experiments that attempt to resolve these
issues on real systems and real channels.  The focus here is on the
following systems
\begin{enumerate}
\item {\bf Land Mobile Wireless} -- Mobility and multipath typical of
this environment will be the focus of the study presented in this paper.
\item {\bf Frequency flat channels} -- A vast majority of the work in
space-time signaling has used frequency flat models.  This corresponds to
relatively narrowband transmission in a traditional land mobile wireless
channels.
\item {\bf Linear Modulations} -- Only linear modulations will be the
focus of the study presented in this paper.
\item {\bf Short Packet Communication} -- The packet lengths of the
system presented in the paper will be approximately 300 symbols.  This
type of system is typical of speech communication systems or short packet
data (paging).
\end{enumerate}
Within this fairly focused area this paper will address:
\begin{enumerate}
\item {\bf Signal Design and Number of Receiver Antennas} -- With a small
number of receive antennas the theory indicates signal design is dominated
by the Hamming distance and the product measure of the pair--wise signal
error matrix.
With a large number of receive antennas the signal design is dominated by
the Euclidean distance of the pair--wise codeword difference.  The
question at what number of receive antennas is the transition between
these two design environments manifested and how significant is the
difference in realistic environments.
\item {\bf Impact of Spatial Correlation} -- Code performance is very
much a function of the spatial correlation between the transmission paths
\cite{SFG:00}.  Consequently it is useful to see if any interesting
characteristics are produced in realistic array geometries that impact the
choice of coded modulations in practice.

\end{enumerate}
This paper is organized with Section II overviewing the models,
Section III
detailing the design paradigms, Section IV presenting the
experimental system, Section V providing the experimental results, and
Section VI concludes.
\section{Signal Models}
For linear modulation the signal at
the $i^{\mbox{\scriptsize th}}$ transmit antenna is modeled as
\begin{equation}
\randx_i(t)=\sum_{l=1}^{N_f} \randx_i(l) u(t-(l-1)T)
\end{equation}
where $u(t)$ is a Nyquist pulse shape and $X_i(l)$ is the modulation
symbol on the $i^{\mbox{\scriptsize th}}$ antenna at symbol $l$ and
$N_f$ is the length of the frame.  If the fading is slow enough, the
sampled matched filter outputs are the sufficient statistics for the
demodulation and the output samples of the matched filter for the
$k^{\mbox{\scriptsize th}}$ symbol are given as a
$L_r \times 1$ vector
\begin{equation}
 \rvy(k)=\rmh(k) \sqrt{E_s}\rvx(k) + \rvn(k)
\end{equation}
where $E_s$ is the energy per transmitted symbol; $\randh_{ij}(k)$
is the complex path gain from transmit antenna $j$ to receive
antenna $i$ at time $kT$; $\rvx(k)$ is the $L_t \times 1$ vector
of symbols transmitted at symbol time $k$; $\rvn(k)$ is the
additive white Gaussian noise vector of size $L_r \times 1$. The
noise is modeled as an independent circularly symmetric zero-mean
complex Gaussian random variable with variance $N_0/2$ per
dimension.

{\bf Coherent demodulation} refers to the case of finding the most likely
transmitted word when the channel is known.  The optimum word
demodulator denoted maximum likelihood (ML) word demodulator. Denote
$\rvb$ as the transmitted word.  For orthogonal modulations when
$\rmh(k)=\mtrxh(k)$ the optimum word demodulator has a simpler form given
as
\begin{eqnarray} \label{eq:cohdemod}
\hat{\rvb}&=&\arg \min_n \sum_{k=1}^{N_f}
\left(\rvy(k)-\vec{s}(n)\right)^H
\left(\rvy(k)-\vec{s}(n)\right)
\end{eqnarray}
where $\vec{s}(n)=\sqrt{E_s}\mtrxh(k)\vx^{(n)}(k)$ is the vector
of noiseless received points on each of the antennas,
$\vx^{(n)}(k)$, a $L_t \times 1$ vector,  is used to denote the
transmitted codeword  at $k^{\mbox{\scriptsize th}}$ symbol for
transmitted bit sequence $n$, $\rvy(k)$ is the received matched
filter output for the $k^{\mbox{\scriptsize th}}$ symbol.  The ML
demodulator essentially finds the transmitted symbol or code
matrix, $\rmx=\mtrxx_n$ that produces the minimum distance between
the matched filter outputs, $\rvy(k)$ and the channel output,
$\sqrt{E_s}\mtrxh(k)\vx^{(n)}(k)$.  If the modulation is defined
on a trellis then the Viterbi algorithm can be used to find this
minimum distance transmitted codeword and if the transmitted
codeword is defined by a lattice then a lattice search algorithm
can be used to find the best codeword.

\section{Overview of Code Design Paradigms}
This section will discuss the different design paradigms for wireless
communications that are often invoked by researchers.  Since this
experiment is focussed on frequency flat MIMO signalling the standard
assumption in this field is that the channel is well modeled by Rayleigh
fading so the brief discussion here will focus on the results for
Rayleigh fading.  Let ${\bf X}$ be the two dimensional code word matrix transmitted by the
space-time modem, and the space-time code ${\cal X}$ be the
collection of these code words.

The code design criteria for {\bf coherent} demodulation in spatially white
Rayleigh fading for systems with a small receive array size are
\cite{Gea:96,TSC:98}
\begin{itemize}
\item
{\em Diversity Advantage\/}:  Maximize
$\Delta_H(n_1,n_2)=\mbox{rank}\left({\mathbf x}_{n_1}-{\mathbf
x}_{n_2}\right)$ over all pairs of code words, ${\mathbf
x}_{n_1} \ne {\mathbf x}_{n_2}$ and ${\mathbf x}_{n_1}, {\mathbf
x}_{n_2}\in{\cal X}$.
\item
{\em Coding Gain\/}:  Maximize the geometric mean of the nonzero
eigenvalues of the signal matrix \\${\bf C}_s=\left({\bf x}_{n_1}-{\bf
x}_{n_2}
\right) \left({\bf x}_{n_1}-{\bf x}_{n_2}\right)^H$ over all
distinct pairs of code words ${\bf x}_{n_1}, \ {\bf x}_{n_2} \in
{\cal X}$.
\end{itemize}
The rank is often denoted the Hamming distance and the geometric mean is
often denoted the product measure to show the relation to single
antenna Rayleigh fading design \cite{DS:88}.  A great deal of work has
gone into designing codes based on these design criteria.

For a large receive array size the design criteria changes to be focussed
more on Euclidean distance \cite{BT:01,AEF:03,CYV:01} and this design
criteria can be stated succinctly as
\begin{itemize}
\item
{\em Euclidean Distance\/}:  Maximize over all distinct pairs of
code words ${\bf x}_{n_1}, \ {\bf x}_{n_2} \in {\cal X}$ the arithmetic
mean of the eigenvalues of ${\bf C}_s=\left ({\bf x}_{n_1}-{\bf
x}_{n_2}\right)\left ({\bf x}_{n_1}-{\bf x}_{n_2}\right)^H$.
\end{itemize}
A reader should note that the boundary between the two scenarios
is not well defined but has been seen in simulation to be around 3 or 4
receive antennas where the channels are modeled as spatially white.
\section{Experimental System}
The experimental system that has been deployed for this experiment is a
narrowband $3 \times 4$ MAR system.  We have chosen a carrier frequency of
220MHz and a bandwidth of around 4kHz.  All modulations are linear
modulation with a spectral raised cosine pulse shape with an excess
bandwidth of 0.2 and a symbol rate of 3.2kHz.  This carrier
frequency and bandwidth allow us to do realistic land mobile testing and
still be confident that the frequency flat assumption will be valid.

\subsection{Radio System}
The UnWiReD narrowband testbed is a software defined
real-time $3 \times 4$ multi-antenna testbed.  The information
bits are encoded and pulse shaped by two Analog Devices
(ADI) fixed point digital signal processors (DSP). The baseband signals
are then digitally up converted to 10MHz IF signals. The 3-TX up
converter radio further up converts the IF signals to the 220MHz RF and
amplifies it for transmission with a maximum transmission power of
35dBm.

%
The receiver chain provides a high performance system for narrowband MIMO
processing.  The received signals are down converted from RF to 10MHz IF
signals by a 4-channel down converter radio and then digitally down
converted to baseband by a 4-channel digital receiver. The 4-channel
digital receiver over--samples the input signals at 64MHz. Overall
receiver dynamic range is greater than 80dB. The overall error vector
magnitude through both the transmit and receive chains is less than
2\%.  The demodulation is performed by two floating point ADI DSPs. The
demodulated data, as well as other important test information, is
transferred to a laptop for data recording and displaying real-time test
results. This data provides a near complete characterization of the
system performance.

\subsection{Packet Format}
The frame for the transmitted signals of this experimental system was
designed to allow many modulations to be tested in a time interleaved
fashion.  This comparison is enabled at the transmitter
by implementing a superframe that is repeated about every 4 seconds.
During this superframe a preamble is sent and 42 different frames of
space-time modulations can be transmitted. The preamble has a signal
format that allows high performance symbol time estimation (a dotting
pattern) so that accurate timing and a course frequency offset can be
acquired.  Each of the subsequent frames or data packets are 300 symbols
in length (93.75ms). Modulations are independent from frame to frame for
the experiments documented in this paper.  At the end of the superframe
there is a silence period of about 70 symbols.  The noise power which can
vary significantly at 220MHz in various scenarios due to man--made noise
is measured every frame and averaged to get a good estimate of the SNR.

\begin{figure}
  \centering
  \includegraphics[width=3.5in]{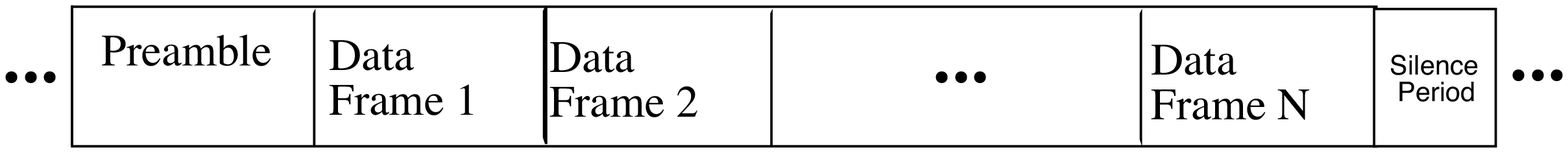}
  \caption{The superframe used in the field experiments.}
  \label{fig:frame}
  \vspace{-0.2in}
\end{figure}

\subsection{Receiver Processing Overview}
All of the receiver functions are implemented in real-time in a
digital signal processor.  Time estimation is derived by using a
nonlinear open loop timing estimator.  Frequency estimation and
frame synchronization are achieved during the first preamble
portion of the frame during the decoding.  Having a unique word
for frame synchronization allows pilot symbols to be inserted and
used for channel estimation and provides block boundary
synchronization for all coded modulations during demodulation. The
details of the pilot symbol processing are given in
Section~\ref{sec:pilot}.  For coherent demodulation ML receivers
based on a trellis search and on a sphere/lattice decoder have
been implemented. This great flexibility allows many algorithms to
be compared to understand the complexity--performance trade-offs
in real implementations. One of the powerful characteristics of
the programmable implementation is that the same transmitted data
can be used to compare decoding with the number of receive
antennas. In almost all modulation formats, decoding for any
number of receive antennas (from $L_r=1$ to $L_r=4$) can be
accomplished in real time. The notable exception to this was the
sphere decoder, real-time sphere/lattice decoding was only able to
be implemented for the $2 \times 2$ case.

\subsection{Channel Estimation} \label{sec:pilot}
Accurate estimation of the channel is crucial for
reliable decoding of coherent coding schemes. Pilot symbol assisted
demodulation (PSAD) is  employed when good performance in
high-mobility situations is desired at a reasonable complexity.  Pilot
symbol based frame design and channel estimation is essentially an
exercise in sampling and optimal interpolation of Gaussian processes
\cite{Cav:91,KF:94,Gea:96}.  Due to the Gaussian nature of the
assumed Rayleigh fading, linear interpolation is optimal.  For a finite
frame size, interpolation at the frame edges performs worse hence it is
important to have more samples at the frame edges. Uniform pilot sampling
in the middle of the frame is optimal as long as the sampling is above the
Nyquist rate of the channel. Guey et al. \cite{Gea:96} showed that
orthogonal pilot elements on each transmit antenna have many desirable
characteristics. An orthogonal pilot symbol pattern maintains good
performance (but not orthogonality at the receiver) even with high
mobility.

\begin{figure}
  \centering
  \includegraphics[width=3.5in]{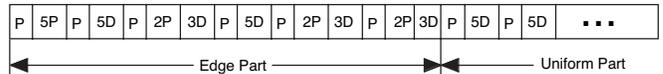}
  \caption{The frame design for the pilot symbol processing for $L_t=2$.}
  \label{fig:psamframe}
  \vspace{-0.1in}
\end{figure}

 The pilot symbol frame structure
for this experiment is optimized for the short frame structure and the
rapid fading that is possible with high mobility. The pilot symbol frame
is optimized separately for different number of transmit antennas. For
example, the 2 Tx frame is shown in Figure \ref{fig:psamframe}. In this
example 72 out of 300 total symbols are used for training.  Hence to
maintain a fair comparison with a modulation/demodulation not needing
training for channel estimation, a code rate increase of roughly 4/3 needs
to be implemented for the coherent coding and decoding.  The  channel gains between
any transmitter-receiver pair are assumed to be  spatially independent for
interpolation filter design. Also, the channel
coefficients are assumed to be constant  over a symbol period but vary
from symbol to symbol according to  Clarke's model \cite{Cla:68} which has
$R_H(m)=J_0(2\pi f_DT m)$ where $J_0$ is the zeroth order Bessel function of the
first kind and
$f_D$ is the Doppler spread of the channel. An FIR Wiener
filter optimized  for
$E_b/N_0 = 30dB$ and Doppler fading rate
$f_DT = 0.01$ is used for  pilot interpolation in the experiments
reported in this paper.

\section{Experimental Results}
\begin{figure}
  \centering{
    \subfigure[Tx deployment.]{\includegraphics[width=1.93in, height=1.5in]{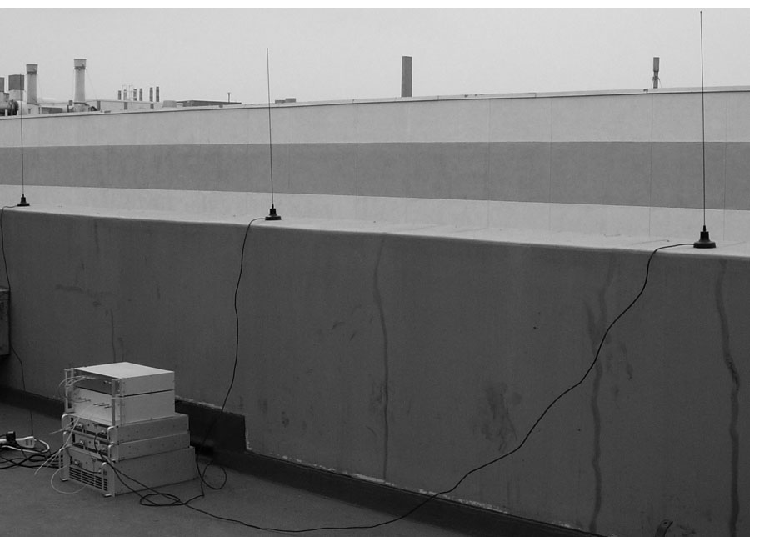}
    \label{fig:txdeploy}}
    \subfigure[Rx deployment.]{\includegraphics[width=1.25in, height=1.5in]{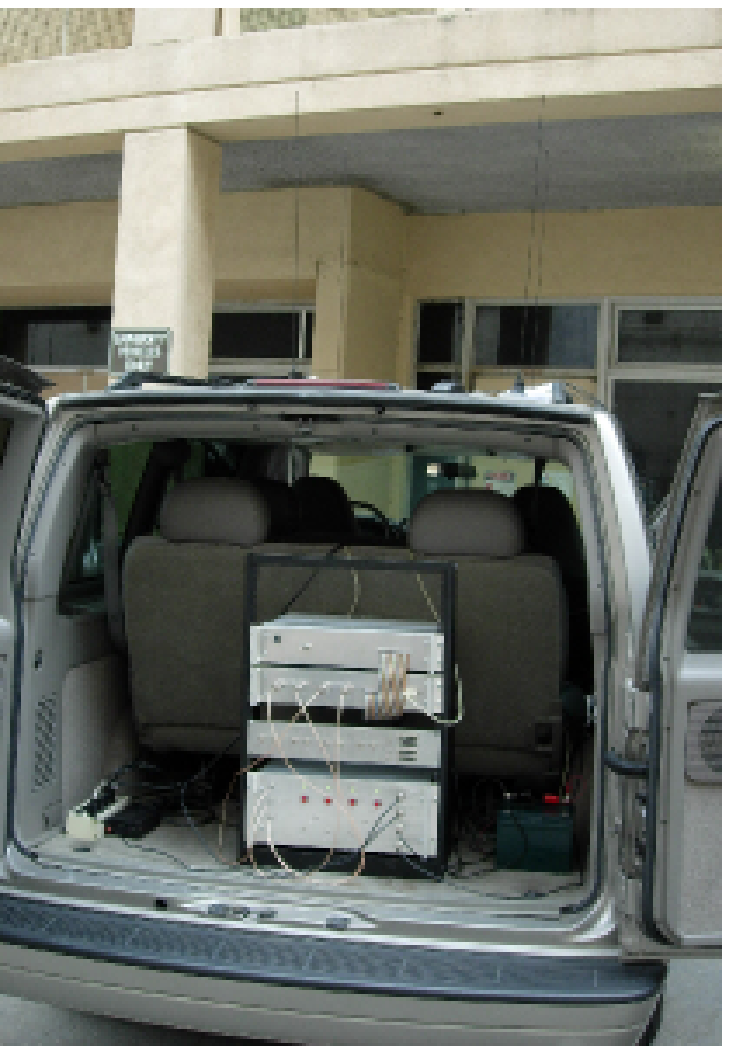}
    \label{fig:rxdeploy}}}
  \caption{Example deployment for the outdoor tests.}
  \label{fig:outdoor}
  \vspace{-0.2in}
\end{figure}

The experimentation was done on the UCLA campus
and the surrounding West Los Angeles area.  The testing reported in
this paper was limited to the scenario where one radio (TX) was deployed
on the top of a 5 story building and one radio (RX) on a vehicle (a cart
or a van).  The test consisted of the receiver radio being driven
around the campus area.  The speed of the driving was maintained at a
rate of less than 5 miles per hour as the codes that were tested were
all designed for quasi-static fading.  The UCLA campus area is heavily urbanized
and a line of sight was not achieved in any significant portion of
the testing. Unless
otherwise specified the receiver array was square with a $\lambda/2$
spacing on each side and the transmitter array was linear with a
$2\lambda$ spacing. An example of the testbed deployment is
shown in Fig.~\ref{fig:outdoor}.

%

The major findings reported
will be the bit error rate and frame error frame versus $E_b/N_0$. The
measured $E_b/N_0$ reported in the experiments are computed by the
averages over the entire superframe and all the receive antennas.  This
measure gives something closer to the average SNR in high mobility tests
and something closer to the instantaneous SNR in static testing but the
measurement was viewed as the best compromise in reporting the data.
The transmitted power from all the antennas in each of the experiments was roughly
15dBm.
\subsection{Code Design and Number of Receive Antennas}
Here a comparison is made of the performance of the various
proposed design methodologies and resulting $R=2$ QPSK $L_{t}=2$ and
$L_{t}=3$ space-time
codes on real channels.  Specifically we would like to understand the
impact of the number of receiver antennas on performance. The codes that
are considered are
\begin{enumerate}
\item 16 state Yan and Blum (YB) \cite{YB:00} is a code optimized for Hamming
distance and product measure.
\item 32 state Chen, Yuan, and Vucetic (CYV) \cite{CYV:01} code is
optimized for Euclidean distance.
\item 32 state superorthogonal code (SO) \cite{SF:02,SF:01} codes were
optimized for simultaneously for Hamming
distance, Euclidean distance, and product measure.
\item 32 state spatially multiplexed traditional (SMT) codes \cite{AEF:03}
were optimized simultaneously for Hamming  distance, Euclidean distance, and
product measure.
\item 32 state universal trellis codes \cite{KW:03} were
optimized to give good performance on any channel that can has a
capacity above $R=2$ bits per channel use.
\end{enumerate}
The general expectation derived from the theory before the experiment was that
codes that were designed for Hamming distance would work well at small number of
receive antennas and codes that are designed for Euclidean distance
would work well with a large number of receive antennas.

\begin{figure}
  \centering
  \includegraphics[width=3.3in]{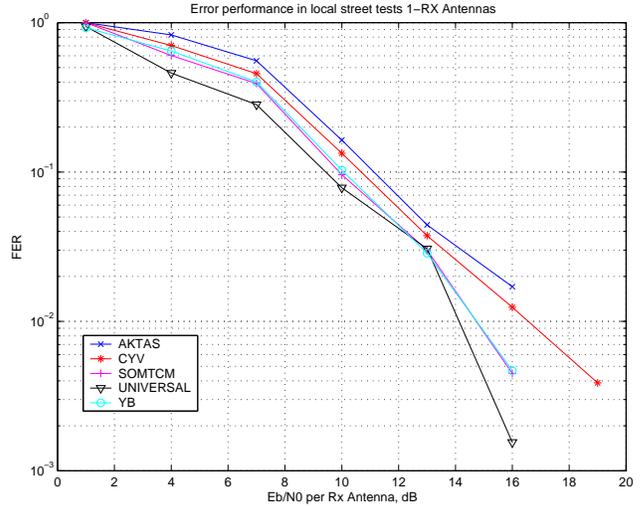}
  \caption{FER performance for $L_{r}=1$.}
  \label{fig:2tx_1rx}
  \vspace{-0.1in}
\end{figure}

\begin{figure}
  \centering
  \includegraphics[width=3.3in]{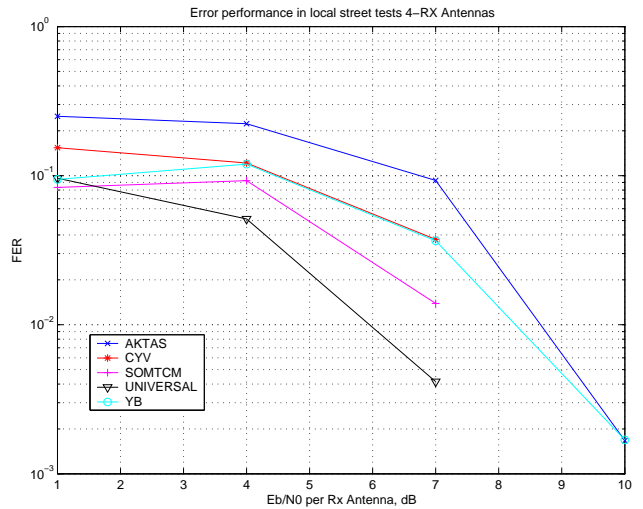}
  \caption{FER performance for $L_{r}=4$.}
  \label{fig:2tx_4rx}
  \vspace{-0.1in}
\end{figure}

A wide variety of data versus measured average
SNR has been compiled for the drive tests. The results of the frame error
rates (FER) for $L_{t}=2$ are summarized in Fig.~\ref{fig:2tx_1rx} and
Fig~\ref{fig:2tx_4rx}. In Fig.~\ref{fig:2tx_1rx} as expected the
designs which have optimized Hamming distance and product measure (YB and SO)
seem to do comparatively
well.  In fact all codes seem to perform very close to the same
level with the universal code having a slight advantage in performance. This is
perhaps not unexpected as this is the only code not designed under the
assumption of spatially white Rayleigh fading in the group tested.
In  Fig.~\ref{fig:2tx_4rx} we observe some unexpected results.  The
universal code shows the best performance followed by the SO code.  Noticeably
worse  performance is observed with both the YB codes and the CYB codes.   This
result is also curious in that these codes have used different  design criteria in
terms of number of receive antenna and yet on real channels seem to produce close
to the  same performance.  Also surprising was the performance of the (SMT)  codes
as they did not seem to be able to use the joint design of both  Euclidean or
Hamming distance to beat codes that optimized either  metric individually (YB or
CYV).  Bit error performance and results for $L_{t}=3$
 coding schemes can be markedly different, but this data is not
reported here due to space constraints.

\subsection{Antenna Spacing Impacts on Code Design}
This section reports on a series of experiments whose goal is to
evaluate the impact of antenna separation on code performance and
code design.  Two types of signalling is considered in this section:
1) multiplexing type schemes (space-time constellations and precoding)
and 2) space-time codes designed to harvest diversity or performance.
Three antenna configurations are considered: 1) 2 $\lambda$ spacing at
the transmitter with 0.5$\lambda$ spacing at the receiver, 2) 2 $\lambda$ spacing at
the transmitter with 0.25$\lambda$ spacing at the receiver, and  $\lambda$ spacing at
the transmitter with 0.25$\lambda$ spacing at the receiver.

\begin{figure}
  \centering
  \includegraphics[width=3.3in]{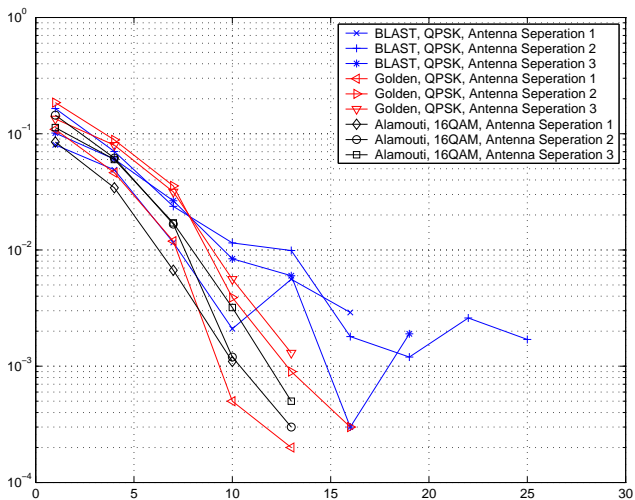}
  \caption{FER performance for $L_{t}=2$ and $L_{r}=2$.}
  \label{fig:constellation}
  \vspace{-0.1in}
\end{figure}

For the case of precoding and space time constellations the Alamouti
\cite{Ala:98} coding scheme has the most robustness to different
antenna geometries.  All constellations considered here use $R=4$ bits per
symbol and the comparison for frame error rate is in
Fig.~\ref{fig:constellation} for $L_{t}=2$ and $L_{r}=2$.  The Alamouti code must use a 16QAM constellation
to achieve this rate while spatial multiplexing and the threaded architectures
can use QPSK constellations.  When the spacing of the array is larger
(closer to spatial independence) the Golden Code \cite{BRV:04} has
the advantage due to the smaller constellation point differences. As
the spacing become closer and the channels become correlated the
Alamouti code has the advantage.  Direct spatial multiplexing
with a BLAST like architecture is clearly lower performance than
either of the two considered architectures and suffers from not achieving full diversity.

\begin{figure}
  \centering
  \includegraphics[width=3.3in]{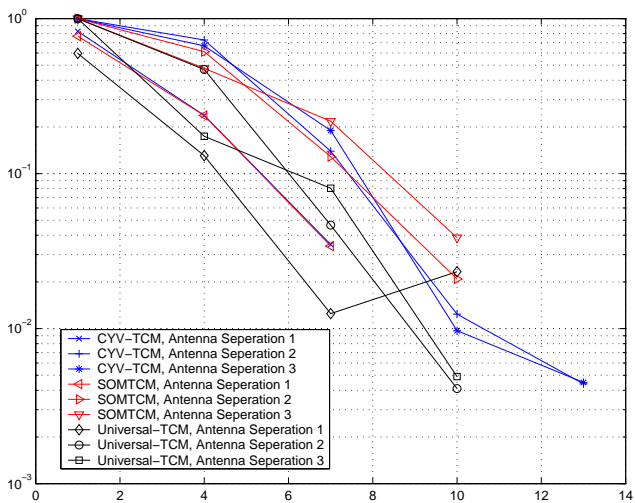}
  \caption{FER performance for $L_{r}=4$.}
  \label{fig:TCMAntsep}
  \vspace{-0.1in}
\end{figure}

For the case of trellis codes, the universal code has the most robustness to different
antenna geometries.  All constellations considered here use a $R=2$ bits per
symbol and the comparison for frame error rate is in
Fig.~\ref{fig:TCMAntsep} for $L_{t}=2$ and $L_{r}=2$.  The universal code has good performance in
most cases with moderate degradation due to spatial correlation.  The
SO and CYV codes show more significant degradation due to spatial
correlation.

\section{Conclusion}
The paper has presented field tests and the conclusions that can be
drawn from the field tests for space--time coding with a variable
number of antennas and varying antenna array size.


\section*{Acknowledgment}
The National Science Foundation has provided
support for this effort which had a time constant of not months but
years.  Analog Devices provided hardware
and software support in the construction of this testbed.




%

\bibliographystyle{IEEEtran}
\bibliography{STCFitz}


\end{document}